\documentstyle[12pt]{article}
\begin{document}
\begin{titlepage}
\begin{centering}
\vspace{5cm}
{\Large\bf Positronium Spectroscopy}\\
\vspace{0.5cm}
{\Large\bf in}\\
\vspace{0.5cm}
{\Large\bf a Magnetic Field}\\
\vspace{2cm}
Jan Govaerts\\
\vspace{0.5cm}
{\em Institut de Physique Nucl\'eaire}
\\
{\em Universit\'e Catholique de Louvain}
\\
{\em 2, Chemin du Cyclotron}
\\
{\em B-1348 Louvain-la-Neuve, Belgium}\\
\vspace{3cm}
\begin{abstract}

\noindent Hyperfine spectroscopy of positronium formed in the
presence of a static magnetic field is considered. Generalising the
situation hitherto developed in the literature, the magnetic field
is not assumed to be parallel to the momentum of incoming
polarised positrons, while the possibility of electron
polarisation is also included in the analysis.
The results are of application to high sensitivity
positron polarimeters used in current $\beta$ decay experiments.

\end{abstract}

\vspace{30pt}

hep-ph/9308247

\end{centering}
\end{titlepage}

\section{Introduction}

Hyperfine structure in positronium is a basic
physical fact with wide ranging
implications. The energy difference between singlet and triplet
states, and their lifetimes dominated by
disintegrations into two and three photons, respectively, are observables
providing crucial tests
for quantum electrodynamics (QED)\cite{Kin} and the
Standard Model\cite{Alc} of electroweak and strong interactions.
On a more practical level, positronium physics is also essential
in a series of technological developments both in the design
of detectors for particle physics experiments and in applications to
problems of solid state physics (for a detailed review
with references to the original literature, see Refs.\cite{Kin,Rich}).

In particular, hyperfine structure of positronium
has been put to use since over thirty years\cite{Dick}
in measurements of positron
polarisations. One instance where such experiments come
immediately to bear on the structure of fundamental interactions
is in the case of positrons emitted in the $\beta$ decay of
radioactive nuclei.
Specifically, any deviation from purely right-handed
polarisation of positrons emitted in the $\beta$ decay of polarised nuclei
would point to physics beyond the Standard Model in its electroweak
sector. Actually, experiments with this purpose in mind and
using such a positronium based positron polarisation measurement
technique, have been pursued at our Institute\cite{F1,F2,M1}.
Results
are promising\cite{F1,F2},
and compete well with limits on physics beyond the
Standard Model obtained from experiments at much higher energies.

Typically in such $\beta$ decay experiments,
one is interested in measuring the longitudinal
polarisation of positrons emitted parallel or antiparallel
to the direction in which the
decaying nucleus is polarised. The emitted positron is stopped
in some specific medium where positronium is formed. The medium being placed
in a strong magnetic field\footnote{Since positrons and electrons
have opposite electric charges, an external {\it electric}
field does not affect positronium states nor their {\it intrinsic} decay
rates (however, ``pick-up" processes of external electrons or other
``quenching" effects can affect these rates\cite{Rich}).
Nevertheless, an external electric field can indeed improve\cite{Pepe}
the positronium formation rate in a given medium.}, the decay
spectrum of positronium formed
provides information\cite{Dick}
on the polarisation of the incoming positron.
The analysis of actual experimental positronium spectra
usually assumes that, at the location where positronium is
formed, the applied magnetic field is {\it exactly} parallel
to the direction in which the positron is emitted.
Since, strictly speaking, such an assumption can never be correct
in practice, the more general situation
has to be considered, namely when the direction of the magnetic field
is arbitrary with respect to the positron momentum.
However, the present author has been unable to find
in the literature\cite{Rich,Dick,Hal,Mad}
a detailed discussion of this point, hence this
note addressing the problem specifically.

In sect.2, a brief review of positronium hyperfine structure in the
absence of any external electromagnetic field is presented, with
also the purpose of specifying our notation. Sect.3
develops the discussion of the positronium ground state
and its hyperfine structure in the presence
of an arbitrary static magnetic field. Sects. 4 and 5 consider
lifetimes and time evolution of hyperfine populations,
respectively, while
sect.6 ends with some conclusions.

\vspace{10pt}

\section{Hyperfine Structure of Positronium}

As is well known, the $1S$ positronium ground state is in fact
split into two hyperfine levels of total spin $S=0$ and $S=1$.
Their energy difference is due to
spin-spin interactions between the positron and the electron,
to relativistic corrections to their kinetic energies
and to virtual pair annihilation in
the $S=1$ channel\footnote{The latter effect is absent in the
$S=0$ channel due to charge conjugation. A photon has $C=-1$,
whereas the $S=0$ and $S=1$ states have $C=+1$ and $C=-1$,
respectively.}.

A non relativistic representation of the
associated wave functions is sufficient for our purposes,
as well as being justified. Accordingly, wave functions
separate into space and spin components, with the space
component being identical for both hyperfine states. Namely,
the $S=0$ singlet or parapositronium
state is given by a wave function of
the form\footnote{Note that a compactified notation
for state vectors is used throughout,
whereby their isotropic radial dependence
is not displayed explicitly.}
\begin{equation}
\mid 0,0>=\psi(r)\ \frac{1}{\sqrt{2}}
\left[\ |+>|-> -\ |->|+>\ \right]\ .
\label{eq:(0,0)}
\end{equation}
Here, $\psi(r)$ is the space wave function of the $1S$ ground state,
while the second factor in the r.h.s. of this expression is the
spin wave function. The convention used throughout is that in the
product of two spin ket vectors, the first always stands for the
electron spin component whereas the second stands for the positron
spin component. These components are taken with respect, say to the
$z$ axis of some reference frame. Later on, this axis will of course
correspond to the direction
in which the $\beta$ decay positron is emitted. It is also assumed
that the space wave function $\psi(r)$ is properly normalised,
\begin{equation}
4\pi\ \int_0^{\infty}dr\ r^2\,\mid\psi(r)\mid ^2=1\ ,
\end{equation}
and that the basis vectors $|+>$ and $|->$ in spin space both
for the electron and for the positron are normalised in the usual way,
namely
\begin{equation}
<+|+>\ =1=\ <-|->\ ,\ \
<+|->\ =0=\ <-|+>\ .
\end{equation}
Consequently, the $\mid 0,0>$ state in (\ref{eq:(0,0)}) is also
of norm $1$.

Similarly,
the three components $m=\pm 1,0$ of the $S=1$ triplet
orthopositronium state are
given by the normalised state vectors
\begin{displaymath}
\mid 1,m=1>=\psi(r)\ |+>|+>\ ,
\end{displaymath}
\begin{equation}
\mid 1,m=-1>=\psi(r)\ |->|->\ ,
\label{eq:(1,m)}
\end{equation}
\begin{displaymath}
\mid 1,m=0>=\psi(r)\ \frac{1}{\sqrt{2}}
\left[\ |+>|->+\ |->|+>\ \right]\ .
\end{displaymath}

The hyperfine states in (\ref{eq:(0,0)}) and
(\ref{eq:(1,m)}) are eigenstates of the total Hamiltonian $H_0$
of the system in the absence of an external magnetic field.
In the non relativistic limit, this Hamiltonian is comprised
of the ordinary Schr\"odinger type
Hamiltonian $H_{(C)}$---which, apart from the usual kinetic
term, only includes
the Coulomb interaction between the two
oppositely charged particles---to which diverse
relativistic corrections are added. The latter contributions
correspond to spin-spin interactions between the electron and
the positron, to relativistic corrections to their kinetic energies,
to virtual pair annihilation effects
for orthopositronium states and to
further QED radiative corrections. Restricting for a moment the discussion
to the non relativistic purely Coulomb Hamiltonian,
the singlet and triplet states above are degenerate eigenstates
of $H_{(C)}$ with energy\footnote{Throughout,
$\alpha$, $m$ and $c$ of course stand for the fine structure constant,
the electron and positron mass and the speed of light, respectively.
Numerical values for these parameters are from Ref.\cite{PDG}.}
\begin{equation}
E_{(C)}=-\frac{1}{4} \alpha^2 m c^2=-6.803\ {\rm eV}\ .
\label{eq:EC}
\end{equation}
In fact, in this purely Coulomb limit, the space wave function
is simply
\begin{equation}
\psi_{(C)}(r)=\left(\pi a^3\right)^{-1/2}\ e^{-r/a}\ ,
\label{eq:wv}
\end{equation}
with the positronium radius
\begin{equation}
a=\frac{2\hbar}{\alpha m c}=1.0584\ {\rm \AA}\ .
\label{eq:a}
\end{equation}
Even though the {\it complete} wave function $\psi(r)$
in (\ref{eq:(0,0)}) departs from
the simple radial dependence in (\ref{eq:wv}),
the parameter $a$ in (\ref{eq:a}) gives a measure of
the spatial extension of the positronium ground state.

Relativistic effects just mentioned lift
the singlet-triplet degeneracy\footnote{In the purely Coulomb
situation of (\ref{eq:EC}),
the singlet-triplet degeneracy is due\cite{YZ} to a dynamical
$SO(4)$ symmetry explicitly broken by
effects now considered. Nevertheless, the $S=1$ triplet
states $m=\pm 1,0$ remain degenerate since,
in the absence of external electromagnetic fields,
the total Hamiltonian $H_0$
is invariant under rotations in space.} according to the
eigenvalues
\begin{equation}
H_0\mid 0,0>=E_0\mid 0,0>\ ,\ \
H_0\mid 1,m=\pm 1,0>=E_1\mid 1,m=\pm 1,0>\ ,
\end{equation}
with energies\cite{Rich}
\begin{equation}
E_0=\left\{-\frac{1}{4}+\alpha^2\left[-\frac{1}{4}
-\frac{5}{64}\right]
+{\cal O}(\alpha^3,\alpha^2\ln\alpha^{-1})\right\}\
\alpha^2m c^2\ ,
\label{eq:E0}
\end{equation}
\begin{equation}
E_1=\left\{-\frac{1}{4}+\alpha^2\left[\frac{1}{12}
-\frac{5}{64}+\frac{1}{4}\right]
+{\cal O}(\alpha^3,\alpha^2\ln\alpha^{-1})\right\}\ \alpha^2m c^2\ ,
\label{eq:E1}
\end{equation}
and the hyperfine difference\cite{CL1,BY}
\begin{equation}
\begin{array}{clcr}
\Delta E=E_1-E_0&=\left[\frac{7}{3}-\frac{\alpha}{\pi}\left(\frac{32}{9}
+2\ln 2\right)+\frac{5}{6}\alpha^2\ln\alpha^{-1}
+{\cal O}(\alpha^2)\right]\ \frac{1}{4}\alpha^4m c^2\\ \\
&=8.41\times 10^{-4}\ {\rm eV}\ .
\end{array}
\label{eq:DE}
\end{equation}

Except for their first term corresponding to the purely Coulomb
contribution (\ref{eq:EC}), the different contributions of order $\alpha^4$
in (\ref{eq:E0}) and (\ref{eq:E1}) are as follows.
Both in $E_0$ and in $E_1$, the first such contribution is
that of the spin-spin interaction energy,
while the second---common to both expressions---is that due to relativistic
corrections to the total kinetic energy. Finally, the third term of order
$\alpha^4$ in $E_1$ follows from virtual pair annihilation in the
orthopositronium triplet state.

To conclude, let us consider positronium lifetimes.
Due to charge conjugation properties,
the singlet state can only decay into an even number of
photons and the triplet states into an odd number (beginning of course
with three photons). Therefore in very good approximation, parapositronium
decays predominantly into two photons and orthopositronium into three
photons, since compared to each of these processes,
the rate for any further photon pair emission is suppressed each time
by an additional power of $\alpha^2$. Hence, only $2\gamma$ and $3\gamma$
decay processes are considered in this note, and
these two channels are assumed to
encompass all possible decay modes of positronium.

The associated lifetimes, including radiative
corrections, have been
computed within QED\cite{CL2,DV}. For the
singlet state, one has the $2\gamma$ decay rate
\begin{equation}
\begin{array}{clcr}
\lambda_S&=\frac{1}{2}\alpha^5\frac{m c^2}{\hbar}\
\left[1-\frac{\alpha}{\pi}\left(5-\frac{\pi^2}{4}\right)
+\frac{2}{3}\alpha^2\ln\alpha^{-1}+{\cal O}(\alpha^2)\right]\\ \\
&=(0.125209\times 10^{-9}\ {\rm s})^{-1}
=7.98665\times 10^9\ {\rm s}^{-1}\ .
\end{array}
\label{eq:LS}
\end{equation}
Similarly for triplet states, their $3\gamma$ decay rate is
\begin{equation}
\begin{array}{clcr}
\lambda_T&=\frac{2}{9\pi}\alpha^6\frac{m c^2}{\hbar}(\pi^2-9)\
\left[1-(10.266\pm0.008)\frac{\alpha}{\pi}
-\frac{1}{3}\alpha^2\ln\alpha^{-1}+{\cal O}(\alpha^2)\right]\\ \\
&=(142.074\times 10^{-9}\ {\rm s})^{-1}
=7.03859\times 10^6\ {\rm s}^{-1}\ .
\end{array}
\label{eq:LT}
\end{equation}
Note the rather large ratio
\begin{equation}
\frac{\lambda_S}{\lambda_T}\ =\ 1134.695\ .
\end{equation}

\vspace{10pt}

\section{Coupling to a Magnetic Field}

Let us now consider positronium states formed in the presence of
some external static magnetic field $\vec{B}=(B_x,B_y,B_z)$. For all
practical purposes, certainly always realised in actual experimental
conditions, it will be assumed that whenever $\vec{B}(\vec{x})$ might
have a non vanishing gradient, this gradient is nevertheless negligible
on the scale of the spatial extension of the positronium state,
namely
\begin{equation}
a\ \frac{\mid \vec{\nabla} B_i\mid}{\mid\vec{B}\mid} \ll 1\ ,
\ \ {\rm for\ all}\ i=x,y,z\ .
\label{eq:grad}
\end{equation}
Here, $a$ is the positronium radius of (\ref{eq:a}).
Effectively, one may then consider the magnetic field $\vec{B}$
not only to be static but also to be
constant, which is thus the situation to be
assumed in the analysis developed in this note. Furthermore,
the field $\vec{B}$ is
not taken to be necessarily parallel to the $z$ axis with respect to
which spin eigenstates were defined in the previous section, since
in practical applications,
the latter axis is often defined by the positron momentum instead.

The presence of the magnetic field $\vec{B}$ induces an additional
interaction term in the total Hamiltonian for the positronium system.
The total Hamiltonian $H$ now includes the previous Hamiltonian $H_0$
with its lowest energy eigenstates
in (\ref{eq:(0,0)}) and (\ref{eq:(1,m)}), to
which the magnetic coupling to magnetic moments is added, namely
\begin{equation}
H_B=-\vec{\mu}_-\cdot\vec{B}-\vec{\mu}_+\cdot\vec{B}\ .
\end{equation}
Here, $\vec{\mu}_{-}$ and $\vec{\mu}_{+}$ are the electron and positron
magnetic moments, respectively. In terms of their spin
operators $\vec{\sigma}_-/2$ and
$\vec{\sigma}_+/2$, respectively, we have\footnote{Here, $\vec{\sigma}$
are the usual Pauli matrices
defining the spin $1/2$ representation of the (double covering $SU(2)$) of
the rotation group $SO(3)$ in three dimensions.}
\begin{equation}
\vec{\mu}_{\pm}=\mp\mu\frac{\vec{\sigma}_{\pm}}{2}\ .
\end{equation}
The magnetic dipole moment $\mu$ is given by
\begin{equation}
\mu=g\ \frac{e\hbar}{2 m}\ ,
\end{equation}
with the gyromagnetic factor\cite{Sch,Som,Pet,Ad,Kin}
\begin{equation}
g=2\left\{1+\frac{\alpha}{2\pi}\ +
\left[\frac{3}{4}\zeta(3)-3\zeta(2)\ln 2
+\frac{1}{2}\zeta(2)+\frac{197}{144}\right]
\left(\frac{\alpha}{\pi}\right)^2\
+{\cal O}\left(\left(\frac{\alpha}{\pi}\right)^3\right)\right\}\ .
\end{equation}
Therefore, the magnetic energy contribution
to the total Hamiltonian reads
\begin{equation}
H_B=\frac{1}{2}\mu (\vec{\sigma}_{-} - \vec{\sigma}_{+})\cdot\vec{B}\ .
\label{eq:HB}
\end{equation}

To complete this list of notations,
it turns out that the parameter setting the physical scale
of magnetic fields in the present system, is the combination
\begin{equation}
\frac{2\mu}{\Delta E}=\frac{1}{3.628575\ {\rm Tesla}}\ ,
\end{equation}
leading to the definition of the {\it positive} quantity
\begin{equation}
x=\frac{2\mu}{\Delta E}\mid\vec{B}\mid\
=\ \frac{\mid\vec{B}\mid}{3.63\ {\rm Tesla}}\ .
\label{eq:x}
\end{equation}
In addition, it proves convenient\footnote{The actual reason
why these definitions are convenient is the fact that
it is the spin $1/2$ representation of the three dimensional
rotation group which appears throughout.} to introduce the following
combinations of the $B_x$ and $B_y$ components of the magnetic
field $\vec{B}$,
\begin{equation}
B_{+}=\frac{B_x+i B_y}{\sqrt{2}}\ ,\ \
B_{-}=\frac{B_x-i B_y}{\sqrt{2}}\ .
\end{equation}

\vspace{10pt}

Given the total Hamiltonian
\begin{equation}
H=H_0+H_B\ ,
\end{equation}
it is now a simple matter to proceed diagonalising it for
its lowest energy states with spherical symmetry, namely
in the sector of positronium $1S$ hyperfine states of
the previous section. First, one easily
finds
\begin{equation}
H\mid 0,0>=E_0\mid 0,0> + \mu B_z\mid 1,0> - \mu B_{-}\mid 1,1>
           +\mu B_{+}\mid 1,-1>\ ,
\end{equation}
\begin{equation}
H\mid 1,0>=E_1\mid 1,0> + \mu B_z\mid 0,0>\ ,
\end{equation}
\begin{equation}
H\mid 1,1>=E_1\mid 1,1> - \mu B_{+}\mid 0,0>\ ,
\end{equation}
\begin{equation}
H\mid 1,-1>=E_1\mid 1,-1> + \mu B_{-}\mid 0,0>\ .
\end{equation}

Given these results, eigenstates of $H$ and their eigenvalues can
be derived after some work. The corresponding four eigenstates are denoted
$\mid \psi_{S^\prime}>$, $\mid \psi_{T^\prime}>$ and $\mid \psi_{\pm}>$.
In the limit of vanishing magnetic field $\vec{B}$,
these states reduce---possibly up to some phase---to the
singlet $\mid 0,0>$, the triplet $\mid 1,0>$ and the
triplet $\mid 1,\pm 1>$ states, respectively, hence the notation.
In particular, the $\mid \psi_{S^\prime}>$ and $\mid \psi_{T^\prime}>$
states are referred to as the ``pseudo-singlet" and ``pseudo-triplet"
states, respectively.

The pseudo-singlet state is given by
\begin{equation}
\begin{array}{clcr}
\mid \psi_{S^\prime}>=&\frac{1}{\sqrt{2}}\
(1+x^2)^{-1/4}(\sqrt{1+x^2}+1)^{-1/2}\times
\left\{(\sqrt{1+x^2}+1)\mid 0,0> -\right.\\ \\
&\left.-\frac{2\mu}{\Delta E} B_z\mid 1,0>
+\frac{2\mu}{\Delta E} B_-\mid 1,1>
-\frac{2\mu}{\Delta E} B_+\mid 1,-1>\right\}\ ,
\end{array}
\label{eq:ps}
\end{equation}
with the eigenvalue
\begin{equation}
E_{S^\prime}=-\frac{1}{2}\Delta E\
\left[\sqrt{1+x^2}-1\right]\ + E_0\ .
\label{eq:Eps}
\end{equation}

The pseudo-triplet state is
\begin{equation}
\begin{array}{clcr}
\mid \psi_{T^\prime}>=&\frac{1}{\sqrt{2}}\
(1+x^2)^{-1/4}(\sqrt{1+x^2}-1)^{-1/2}\times
\left\{(\sqrt{1+x^2}-1)\mid 0,0> +\right.\\ \\
&\left.+\frac{2\mu}{\Delta E} B_z\mid 1,0>
-\frac{2\mu}{\Delta E} B_-\mid 1,1>
+\frac{2\mu}{\Delta E} B_+\mid 1,-1>\right\}\ ,
\end{array}
\label{eq:pt}
\end{equation}
with the eigenvalue
\begin{equation}
E_{T^\prime}=+\frac{1}{2}\Delta E\
\left[\sqrt{1+x^2}+1\right]\ +E_0\ .
\label{eq:Ept}
\end{equation}

Finally, the remaining two states are
\begin{equation}
\begin{array}{clcr}
\mid \psi_\pm>=&\frac{1}{\sqrt{2}}\
\frac{\sqrt{B^2_x+B^2_y}}{\mid \vec{B}\mid}\ \mid 1,0> +\\ \\
&+\frac{1}{\sqrt{2}}\ \frac{B_z\pm\mid\vec{B}\mid}{\mid\vec{B}\mid}\
\frac{B_-}{\sqrt{B^2_x+B^2_y}}\ \mid 1,1>
+\frac{1}{\sqrt{2}}\ \frac{-B_z\pm\mid\vec{B}\mid}{\mid\vec{B}\mid}\
\frac{B_+}{\sqrt{B^2_x+B^2_y}}\mid 1,-1>\ ,
\end{array}
\label{eq:pm}
\end{equation}
with the degenerate eigenvalue
\begin{equation}
E_\pm=\Delta E + E_0 = E_1\ .
\label{eq:Epm}
\end{equation}
Here, upper (resp. lower) signs in the r.h.s. of (\ref{eq:pm})
correspond to the state $\mid \psi_+>$ (resp. $\mid \psi_->$).

By construction, the four states just given not only diagonalise the
total Hamiltonian $H$, but are also orthonormalised. Namely, these
states are
orthogonal by pairs and are normalised to $1$.
Of course, orthogonality
is automatic for non degenerate states but not for degenerate ones;
by construction, the states $\mid \psi_+>$ and $\mid \psi_->$
given above do indeed have a vanishing inner product.

Before commenting on these expressions, it is useful to consider
the limit in which the magnetic field $\vec{B}$ is parallel to the
$z$ axis, namely when $B_x=0$ and $B_y=0$. Of course, in this
limit, the eigenvalues of the four states above remain unchanged,
since their energies only depend on the variable $x$, {\it i.e.} the
norm of the magnetic field. However, the states specified above,
which diagonalise the
total Hamiltonian $H$, then reduce to
\begin{equation}
\mid \psi_{S^\prime}>=\cos\theta \mid 0,0>
-\sin\theta \mid 1,0>\ ,
\label{eq:ps0}
\end{equation}
\begin{equation}
\mid \psi_{T^\prime}>=\eta\ \left\{\ \sin\theta \mid 0,0>
+\cos\theta \mid 1,0>\ \right\}\ ,
\label{eq:pt0}
\end{equation}
\begin{equation}
\mid \psi_\pm>=\pm e^{\pm i\eta\omega}\ \mid 1,\pm\eta>\ .
\label{eq:pm0}
\end{equation}
In these expressions, $\eta={\rm sign}(B_z)$ is the
sign of $B_z$, $\omega$ is some
arbitrary phase whose value is
dependent on the manner in which the limit
$B_x=0,\ B_y=0$ is taken\footnote{Indeed, the coefficients of
the states $\mid 1,1>$ and $\mid 1,-1>$ in (\ref{eq:pm}) are
non analytic functions of $B_x$ and $B_y$.}, and the mixing angle
$\theta$ is defined by
\begin{equation}
\cos\theta=\frac{1}{\sqrt{2}}\ \sqrt{1+\frac{1}{\sqrt{1+x^2}}}\ ,\ \
\sin\theta=\eta\ \frac{1}{\sqrt{2}}\
\sqrt{1-\frac{1}{\sqrt{1+x^2}}}\ .
\label{eq:theta}
\end{equation}
Of course, these expressions coincide with those usually
found in the literature\cite{Dick,Mad,Rich},
in which case it is customary to take
$\mid \psi_\pm>=\mid 1,\pm 1>$ and
the $z$ axis along the magnetic field $\vec{B}$,
namely $\eta=+1$. In particular, note that in the limit of an
{\it infinite} magnetic field $\vec{B}$, the pseudo-singlet and
pseudo-triplet states in (\ref{eq:ps0}) and (\ref{eq:pt0})
further reduce to
(in the notation of (\ref{eq:(0,0)})\/)
\begin{equation}
\mid \psi_{S^\prime}>=-\eta\ \psi(r) \mid -\eta>\mid +\eta>\ ,\ \
\mid \psi_{T^\prime}>=+\eta\ \psi(r) \mid +\eta>\mid -\eta>\ .
\end{equation}

\vspace{10pt}

In order to comment on the physical significance of these results,
let us first consider the case where the magnetic field is {\it parallel}
to the $z$ axis. Even though the vector $\vec{B}$ then explicitly breaks
rotational invariance of the positronium system in vacuum,
there still remains the symmetry of arbitrary rotations around
the $z$ axis. Consequently, the spin projection $m$
on that axis still defines
a good quantum number for positronium states (of vanishing
angular momentum.). Therefore, the states $\mid 1,1>$ and
$\mid 1,-1>$ must remain eigenstates of the total Hamiltonian,
whereas the other two states with $m=0$, namely
$\mid 0,0>$ and $\mid 1,0>$, are now allowed to mix. Moreover,
for the former two states with $m=\pm 1$, since the electron
and positron have their spins then aligned and since their
magnetic moments are equal in norm but opposite,
the magnetic coupling energy $H_B$ vanishes
identically, leading for these two states
to the same energy eigenvalue $E_1$ as in the
absence of any field $\vec{B}$. Finally, for the remaining two states
with $m=0$ diagonalising the total Hamiltonian $H$, in the
limit\footnote{This situation was pointed out to the author
by J. Deutsch.} where the magnetic coupling $H_B$
is much larger than all other
contributions to $H$, namely for magnetic fields whose magnitude is
much larger than $(\Delta E/(2\mu)=3.63\ {\rm Tesla})$, the state
of lowest (resp. highest) energy, {\it i.e.} $\mid \psi_{S^\prime}>$
(resp. $\mid \psi_{T^\prime}>$), is the one for which the electron
and positron magnetic moments are
both aligned parallel (resp. antiparallel)
to the magnetic field $\vec{B}$, or equivalently
the one for which the electron spin is
aligned antiparallel (resp. parallel) to $\vec{B}$ and the
positron spin is parallel (resp. antiparallel) to $\vec{B}$.

These properties---indeed, solely expected on
physical grounds independently of any explicit calculation---are
beautifully confirmed by the results above in
the case where $B_x=0$ and $B_y=0$. At this stage, it thus
appears that an explicit
calculation serves the purpose only of determining the mixing
angle $\theta$ in (\ref{eq:theta}) between the two $m=0$ states,
and of deriving the energy eigenvalues $E_{S^\prime}$ and
$E_{T^\prime}$ in (\ref{eq:Eps}) and (\ref{eq:Ept}),
as functions of the magnetic
field $\vec{B}=(0,0,B_z)$.

Consider now the general situation when the direction
of the magnetic field $\vec{B}$ is arbitrary with respect to
the $z$ axis. In fact, the results just discussed can be used
in order to understand the general case as well. Indeed, the
only difference between the two configurations is that the axis
with respect to which the spin part of state vectors is
expanded, is different. Hence, by an appropriate change of
basis {\it in the spin sector}, effected through a rotation
in the spin $1/2$ representation,
the eigenstates of the total Hamiltonian $H$ in the arbitrary case
$(B_x,B_y)\neq (0,0)$ can in principle be constructed from
the expressions of these states when $(B_x,B_y)=(0,0)$.
Therefore, since under such a rotation in spin space
representations of spin $0$ and of spin $1$ are invariant, only
the states $\mid 1,m=0,\pm 1>$ can mix among themselves.
Consequently, in the general case, both the pseudo-singlet
and pseudo-triplet states should be given as some
superposition of all {\it four}
states $\mid 0,0>$ and $\mid 1,m=0,\pm 1>$,
with in particular the coefficient of the $\mid 0,0>$ component
{\it independent} of the {\it components} of
the magnetic field but only dependent
on its norm $\mid\vec{B}\mid$,
whereas the remaining two states $\mid \psi_\pm>$ can only
involve the three states $\mid 1,m=0,\pm 1>$.
In addition, for all four eigenstates, the coefficients of the
states $\mid 1,m=0,\pm 1>$ must depend on {\it all three components}
$B_x$, $B_y$ {\it and} $B_z$ of the magnetic field.

Indeed, these are
features of the results in (\ref{eq:ps}), (\ref{eq:pt})
and (\ref{eq:pm}), which therefore find their origin in the fact
that spin $0$ and spin $1$ representations are invariant under
space rotations. However, only an explicit calculation---either
along the lines just sketched or by direct diagonalisation
of the Hamiltonian as
done in this note---can determine the specific mixing coefficients
defining each of the eigenstates of the total Hamiltonian $H$.
Incidentally, note that the same argument of invariance
under space rotations explains why eigenvalues of $H$ remain
unchanged when the magnetic field $\vec{B}$ is no longer
parallel to the $z$ axis, {\it i.e.} why these eigenvalues
only depend on the norm $\mid\vec{B}\mid$ of the magnetic field.

\vspace{10pt}

\section{Positronium Lifetimes}

As a first application of results so far, let us compute
decay rates for all four eigenstates of the total
Hamiltonian $H$ in the presence of a magnetic field $\vec{B}$.
Actually, such a calculation is rather straightforward, given the
decay rates $\lambda_S$ and $\lambda_T$ of the singlet and triplet
states $\mid 0,0>$ and $\mid 1,m=0,\pm 1>$, respectively.

First, consider decay rates into two photons. Due to charge
conjugation, only the $\mid 0,0>$ state has $2\gamma$ decays.
Therefore, the $2\gamma$ decay rate of the pseudo-singlet
state $\mid \psi_{S^\prime}>$ is
\begin{equation}
\lambda^{(2\gamma)}_{S^\prime}=
\ \frac{1}{2}\ \left[1+\frac{1}{\sqrt{1+x^2}}\right]\ \lambda_S\
=\ \lambda_S\cos^2\theta\ .
\label{eq:LSp2}
\end{equation}
For the pseudo-triplet state $\mid \psi_{T^\prime}>$, we have
\begin{equation}
\lambda^{(2\gamma)}_{T^\prime}=
\ \frac{1}{2}\ \left[1-\frac{1}{\sqrt{1+x^2}}\right]\ \lambda_S\
=\ \lambda_S\sin^2\theta\ .
\label{eq:LTp2}
\end{equation}
Finally, the $2\gamma$ decay rate of the remaining two states
$\mid \psi_\pm>$ is
\begin{equation}
\lambda^{(2\gamma)}_\pm=0\ .
\label{eq:Lpm2}
\end{equation}

Similarly, for the $3\gamma$ decay rates of these states in the
same order, one finds
\begin{equation}
\lambda^{(3\gamma)}_{S^\prime}=
\ \frac{1}{2}\ \left[1-\frac{1}{\sqrt{1+x^2}}\right] \lambda_T\
=\ \lambda_T\sin^2\theta\ ,
\label{eq:LSp3}
\end{equation}
\begin{equation}
\lambda^{(3\gamma)}_{T^\prime}=
\ \frac{1}{2}\ \left[1+\frac{1}{\sqrt{1+x^2}}\right] \ \lambda_T\
=\ \lambda_T\cos^2\theta\ ,
\label{eq:LTp3}
\end{equation}
\begin{equation}
\lambda^{(3\gamma)}_\pm=\ \lambda_T\ .
\label{eq:Lpm3}
\end{equation}
In these expressions, the angle $\theta$ is defined in
(\ref{eq:theta}).

Finally, total decay rates---ignoring the much suppressed
rates into four or more photons---are simply
\begin{equation}
\lambda_{S^\prime}=
\ \frac{1}{2}(\lambda_S+\lambda_T)
+\frac{1}{2}\frac{1}{\sqrt{1+x^2}}(\lambda_S-\lambda_T)\
=\ \lambda_S\cos^2\theta + \lambda_T\sin^2\theta\ ,
\label{eq:LSp}
\end{equation}
\begin{equation}
\lambda_{T^\prime}=
\ \frac{1}{2}(\lambda_S+\lambda_T)
-\frac{1}{2}\frac{1}{\sqrt{1+x^2}}(\lambda_S-\lambda_T)\
=\ \lambda_S\sin^2\theta + \lambda_T\cos^2\theta\ ,
\label{eq:LTp}
\end{equation}
\begin{equation}
\lambda_\pm=\lambda_T\ .
\label{eq:Lpm}
\end{equation}

As ought to be expected, these expressions only
depend on the norm of the magnetic field $\vec{B}$, but
not on its direction. Indeed, since the choice of axis with respect to
which spin states are expanded does not affect the calculation
of decay rates\footnote{Indeed, positronium states can
``remember" the direction and polarisation of the incoming positron only
through the {\it populations} of the four Hamiltonian eigenstates (this
is the topic of the next section).
Decay rates are {\it intrinsic} properties of each of these states,
and as such, are thus independent of any variable
possibly affecting positronium formation.},
that axis can always be taken along the
magnetic field, in which case decay rates can only depend on
$\mid B_z\mid$, namely the norm of the magnetic field.
Incidentally, note that {\it total} $2\gamma$ and $3\gamma$
decay rates---$\lambda_S$ and $\lambda_T$, respectively---are
independent of $\vec{B}$---a consequence of unitarity.

It is also easy to check that the statistical decay rates into
two and three photons, {\it i.e.} the average of each of these rates
over the four Hamiltonian eigenstates, are independent of
the magnetic field.
These statistical rates
thus coincide with their values when $\vec{B}=\vec{0}$,
namely $\lambda_S/4$ and $3\lambda_T/4$ for two
and three photon decays, respectively.

Finally, let us remark that the difference between the total
decay rates for the pseudo-triplet and $\mid \psi_\pm>$ states,
\begin{equation}
\lambda_{T^\prime}-\lambda_T=
\ \frac{1}{2}(\lambda_S-\lambda_T)
\ \left[1-\frac{1}{\sqrt{1+x^2}}\right]\
=\ (\lambda_S - \lambda_T)\sin^2\theta\ ,
\label{eq:ldiff}
\end{equation}
is a quantity always positive for all values of the magnetic field.

\vspace{10pt}

\section{Positronium Populations}

Let us now address the specific topic of this note; the
formation of positronium in a medium placed in
a magnetic field. In view of applications, the incoming positron
is assumed to have a polarisation $P_+$ ($-1 \leq P_+ \leq +1$).
This polarisation $P_+$ is the expectation value
of the positron spin projected onto its momentum, the latter
vector thus also defining the $z$ axis
for spin quantisation from now on.
Therefore, up to a physically irrelevant overall phase,
the spin component of the
incoming positron wave function is given by
\begin{equation}
\frac{1}{\sqrt{2}}\sqrt{1+P_+}\mid +>
+\ e^{i\phi_+}\frac{1}{\sqrt{2}}\sqrt{1-P_+}\mid ->\ .
\label{eq:P+}
\end{equation}
Here, $\phi_+$ is an arbitrary {\it phase difference}---thus
possibly leading to observable physical effects---between the two
spin components defining a positron state of polarisation $P_+$.

Similarly, it will be assumed\footnote{The interest of this
possibility was pointed out to the author by F. Gimeno-Nogues.}
that the positron capturing electron
has a polarisation $P_-$ ($-1\leq P_-\leq +1$)
along the same $z$ axis. Though in most
practical applications, the positronium formation medium is at
temperatures such that electrons are effectively not polarised, some
experiments at very low temperatures are being planned,
for which an
investigation of possible effects due to electron polarisation
in the applied magnetic field
might therefore be found useful. Consequently,
again up to a physically irrelevant overall phase, the spin part of
the electron wave function is also of the form
\begin{equation}
\frac{1}{\sqrt{2}}\sqrt{1+P_-}\mid +>
+\ e^{i\phi_-}\frac{1}{\sqrt{2}}\sqrt{1-P_-}\mid ->\ ,
\label{eq:P-}
\end{equation}
where $\phi_-$ is another arbitrary phase shift.

Hence, at the moment of positronium formation, it is assumed that
the spin component of the positronium state vector
$\mid \psi, t=0>$ is simply given
by the tensor product\footnote{Note that it is always possible
to ``rotate away" {\it one} of the two phases $\phi_-$ or
$\phi_+$---{\it but not both}---by an appropriate rotation
around the $z$ axis. However, dependence on the cancelled
phase then reappears through the $B_x$ and $B_y$ components of
the rotated magnetic field.}
of the electron and positron spin vectors
in (\ref{eq:P+}) and (\ref{eq:P-}), while the
space part of the state vector is of course the wave function
$\psi(r)$ in (\ref{eq:(0,0)}). In order to obtain
the time evolution of the associated state vector,
and thus also the time dependence of its decay products,
the resulting wave function $\mid \psi, t=0>$ of formed positronium
has to be expanded in the basis of eigenstates of the
total Hamiltonian $H$ in the presence of the magnetic field $\vec{B}$.
This change of basis thus defines coefficients $C_{S^\prime}$,
$C_{T^\prime}$ and $C_\pm$ such that
\begin{equation}
\mid \psi, t=0>=\sum_{a=S^\prime,T^\prime,+,-}
C_a\ \mid \psi_a>\ .
\end{equation}
Explicit expressions for these coefficients are given
in Appendix A.
Time evolution of the positronium state formed is then given by
\begin{equation}
\mid \psi, t>=\sum_{a=S^\prime,T^\prime,+,-}
C_a\ \mid \psi_a>\
\exp\left(-\frac{i}{\hbar}E_a t
-\frac{1}{2}\lambda_a t\right)\ ,
\label{eq:Pst}
\end{equation}
with the quantities $E_a$ and $\lambda_a$
($a=S^\prime,T^\prime,+,-$) defined in (\ref{eq:Eps}),
(\ref{eq:Ept}), (\ref{eq:Epm}) and (\ref{eq:LSp}), (\ref{eq:LTp})
and (\ref{eq:Lpm}).

Given these expressions, time evolution of each of the populations
associated to each of the four states $\mid \psi_a>$
($a=S^\prime,T^\prime,+,-$) is simply obtained as
\begin{equation}
\mid C_a \mid^2\ e^{-\lambda_a t}\ ,
\ \ a=S^\prime,T^\prime,+,-\ .
\label{eq:popt}
\end{equation}
Expressions for all observables of interest are then easily written
down. For example, $2\gamma$ and $3\gamma$ production rates
are\footnote{Note that when actual experimental data are
considered, the $2\gamma$ production rate in (\ref{eq:R2})
should also include a ``fast" component due to {\it direct}
pair annihilation of incoming positrons with
electrons of the positronium forming medium.}, respectively,
\begin{equation}
R_{(2\gamma)}(t)=\sum_{a=S^\prime,T^\prime,+,-}
\ \lambda^{(2\gamma)}_a\mid C_a\mid^2\
e^{-\lambda_a t}\ ,
\label{eq:R2}
\end{equation}
\begin{equation}
R_{(3\gamma)}(t)=\sum_{a=S^\prime,T^\prime,+,-}
\ \lambda^{(3\gamma)}_a\mid C_a\mid^2\
e^{-\lambda_a t}\ ,
\label{eq:R3}
\end{equation}
while the total photon production
rate---simply the sum of the latter two rates---is itself
\begin{equation}
R(t)=\sum_{a=S^\prime,T^\prime,+,-}
\ \lambda_a\mid C_a\mid^2\
e^{-\lambda_a t}\ .
\label{eq:R}
\end{equation}

The rather lengthy expressions for the populations at $t=0$,
namely the coefficients $\mid C_a\mid^2$
($a=S^\prime,T^\prime,+,-$), are given in Appendix B.
Results probably more relevant at this point are the same
coefficients {\it averaged\/}\footnote{This average
{\it does not} amount to setting $e^{i\phi_\pm}=0$ in the
original expressions for the coefficients $C_a$,
and can only be applied once the {\it complete} expressions
of Appendix B for the
coefficients $\mid C_a\mid^2$ have been obtained.}
over the phase shifts $\phi_-$
and $\phi_+$. Indeed, one ought to expect that for
most sources where the $\beta$ decay process
responsible for positron production is taking place,
positron states with all possible phase shifts $\phi_+$ are
statistically populated, thereby justifying an average of the
final positronium populations over $\phi_+$.
Similarly, in the positronium
forming medium, one should also expect that all electron states
with different phase shifts $\phi_-$ are statistically populated,
again justifying an average over the phase $\phi_-$.
Under these assumptions\footnote{Note that
taking such averages is even more justifiable
in the instance---often realised in practice---that
the magnetic fields present in an experimental set-up possess
an axial symmetry along the axis of incoming positrons.
Indeed, as was noticed previously, either of the phases $\phi_-$
or $\phi_+$ can always be ``rotated away" by an appropriate rotation
around the $z$ axis, then also a symmetry
transformation of the magnetic fields.},
the averaged populations are
given by
\begin{displaymath}
\overline{\mid C_{S^\prime,T^\prime}\mid^2}=
\ \frac{1}{4}\left[1-P_- P_+\right]
\mp\ \frac{1}{4}\frac{1}{\sqrt{1+x^2}}\frac{2\mu}{\Delta E}B_z
\left[P_- - P_+\right]+
\end{displaymath}
\begin{equation}
+\ \frac{1}{2}\left[1\mp\frac{1}{\sqrt{1+x^2}}\right]
\frac{B_- B_+}{\mid\vec{B}\mid^2}P_- P_+\ ,
\end{equation}
and
\begin{equation}
\overline{\mid C_\pm\mid^2}=
\ \frac{1}{4}\left[1\pm\frac{B_z}{\mid\vec{B}\mid}P_-\right]
\left[1\pm\frac{B_z}{\mid\vec{B}\mid}P_+\right]\ .
\end{equation}
In the r.h.s. of these two expressions, the upper (resp. lower)
sign refers to $\overline{\mid C_{S^\prime}\mid^2}$
(resp. $\overline{\mid C_{T^\prime}\mid^2}$) and
$\overline{\mid C_+\mid^2}$ (resp.
$\overline{\mid C_-\mid^2}$), respectively.
Note that the sum of the four phase averaged populations does
indeed reduce to $1$, as ought be the case since,
by definition, the state
$\mid \psi, t=0>$ is normalised to $1$.

In turn, phase averaged photon production rates
$\overline{R}_{(2\gamma)}(t)$, $\overline{R}_{(3\gamma)}(t)$
and $\overline{R}(t)$
are defined as in
(\ref{eq:R2}), (\ref{eq:R3}) and (\ref{eq:R}), of course involving now
the phase averaged coefficients
$\overline{\mid C_a\mid^2}$ ($a=S^\prime,T^\prime,+,-$)
just given. Clearly, these averaged photon rates are more readily
amenable to experimental measurement than are the non phase averaged
rates considered previously.

\vspace{10pt}

\section{Conclusions}

This note reports on the calculation of hyperfine
positronium populations
formed in the presence of an arbitrary magnetic field---whose
gradient is assumed to be vanishingly small over the spatial
extension of the positronium bound state---for arbitrary positron
and electron polarisations. The analysis generalises results
available\cite{Dick,Mad,Rich}
in the literature in two respects. On the one hand,
the magnetic field is {\it not}
assumed to be necessarily aligned along the momentum of the
incoming positron. On the other hand,
allowing for possible electron polarisation
effects enables the present results
to be of application to positron
polarimeters operated at very low temperatures.

Expressions derived in this note provide the basic information
required in any experimental analysis of results obtained using positron
polarimeters based on either time or energy distributions
of positronium decay photons. As a simple but explicit
illustration of relevance
to current $\beta$ decay experiments\cite{F1,F2,M1},
let us consider for example the phase averaged
photon production time spectrum $\overline{R}(t)$
when electrons in the
positronium formation medium are not polarised ($P_-=0$).
Expressions derived in the previous section then lead to
\begin{displaymath}
\overline{R}(t)=
\ \frac{1}{2}\lambda_T e^{-\lambda_T t}
+\ \frac{1}{4}\lambda_{T^\prime}
\left[1-\frac{1}{\sqrt{1+x^2}}\frac{2\mu}{\Delta E}
B_z P_+\right] e^{-\lambda_{T^\prime} t}+
\end{displaymath}
\begin{equation}
+\ \frac{1}{4}\lambda_{S^\prime}
\left[1+\frac{1}{\sqrt{1+x^2}}\frac{2\mu}{\Delta E}
B_z P_+\right] e^{-\lambda_{S^\prime} t}\ .
\label{eq:aveR}
\end{equation}
However, since the
pseudo-singlet decay rate $\lambda_{S^\prime}$ is
much larger than the
pseudo-triplet one $\lambda_{T^\prime}$,
the pseudo-singlet contribution in the r.h.s. of
(\ref{eq:aveR}) becomes effectively negligible
after a nanosecond or so, leaving only the first two terms.
Hence in effect, the photon time spectrum $\overline{R}(t)$
\begin{equation}
\overline{R}(t\geq 1\ {\rm ns})\approx
\ \frac{1}{2}\lambda_T e^{-\lambda_T t}+
\ \frac{1}{4}\lambda_{T^\prime}
\left[1-\epsilon P_+\right] e^{-\lambda_{T^\prime} t}\ ,
\end{equation}
with the analysing power
\begin{equation}
\epsilon=\ \frac{1}{\sqrt{1+x^2}}\frac{2\mu}{\Delta E} B_z\ ,
\end{equation}
provides the means of measuring positron polarisations.
Note that, when compared to the usual
result obtained for a magnetic field $\vec{B}$
assumed to be parallel
to the incoming positron momentum, the sole effect of non
vanishing components $B_x$ and $B_y$ in this simple example
is to {\it decrease} the effective
analysing power $\epsilon$  multiplying
the positron polarisation $P_+$.
Nevertheless, there certainly exist other instances where the
effects of non vanishing components $B_x$ and $B_y$
have to be properly accounted for
when analysing actual experimental data. The results
of this note provide the basis for such
investigations.

\vspace{10pt}

\section*{Acknowledgements}

Useful discussions with J. Deutsch
and F. Gimeno-Nogues
are gratefully acknowledged.

\clearpage
\newpage

\section*{Appendix A}

This Appendix gives the coefficients of the linear combination of the
total Hamiltonian eigenstates defining the state vector of
positronium formed in the presence of a magnetic field
$\vec{B}$ (see (\ref{eq:Pst})\/).

For the pseudo-singlet and pseudo-triplet states, one has
\begin{displaymath}
C_{S^\prime,T^\prime}=
\ \frac{1}{\sqrt{2}}\ (1+x^2)^{-1/4}\ (\sqrt{1+x^2}\pm 1)^{-1/2}\times
\end{displaymath}
\begin{displaymath}
\times\left\{\ \frac{1}{2\sqrt{2}}
\left[\sqrt{1+P_-}\ \sqrt{1-P_+}\ e^{i\phi_+}
-\sqrt{1-P_-}\ \sqrt{1+P_+}\ e^{i\phi_-}\right]
(\sqrt{1+x^2}\pm1)\ \mp\right.
\end{displaymath}
\begin{displaymath}
\mp\ \frac{1}{2\sqrt{2}}
\left[\sqrt{1+P_-}\ \sqrt{1-P_+}\ e^{i\phi_+}
+\sqrt{1-P_-}\ \sqrt{1+P_+}\ e^{i\phi_-}\right]
\frac{2\mu}{\Delta E}B_z\ \pm
\end{displaymath}
\begin{equation}
\pm\ \frac{1}{2}\sqrt{1+P_-}\ \sqrt{1+P_+}\ \frac{2\mu}
{\Delta E}B_+\ \mp
\end{equation}
\begin{displaymath}
\left.\mp\ \frac{1}{2}\sqrt{1-P_-}\ \sqrt{1-P_+}\ e^{i(\phi_-+\phi_+)}
\ \frac{2\mu}{\Delta E}B_-\right\}\ ,
\end{displaymath}

\vspace{10pt}

\noindent where the
upper (resp. lower) sign refers to the coefficient
$C_{S^\prime}$ (resp. $C_{T^\prime}$).

Similarly, the coefficients of the remaining two states
$\mid \psi_\pm>$ are
\begin{displaymath}
C_\pm=\ \frac{1}{4}\
\left[\sqrt{1+P_-}\ \sqrt{1-P_+}\ e^{i\phi_+}
+\sqrt{1-P_-}\sqrt{1+P_+}\ e^{i\phi_-}\right]
\frac{\sqrt{B^2_x+B^2_y}}{\mid\vec{B}\mid}\ +
\end{displaymath}
\begin{equation}
+\ \frac{1}{2\sqrt{2}}\sqrt{1+P_-}\ \sqrt{1+P_+}
\ \frac{B_z\pm\mid\vec{B}\mid}{\mid\vec{B}\mid}
\frac{B_+}{\sqrt{B^2_x+B^2_y}}\ +
\end{equation}
\begin{displaymath}
+\ \frac{1}{2\sqrt{2}}\sqrt{1-P_-}\ \sqrt{1-P_+}\ e^{i(\phi_-+\phi_+)}
\ \frac{-B_z\pm\mid\vec{B}\mid}{\mid\vec{B}\mid}
\frac{B_-}{\sqrt{B^2_x+B^2_y}}\ .
\end{displaymath}

\vspace{10pt}

\noindent Here again,
the upper (resp. lower) sign refers to the
coefficient $C_+$ (resp. $C_-$).

\clearpage
\newpage

\section*{Appendix B}

This Appendix gives the populations of positronium states
formed in the presence of a magnetic field, for electrons
and positrons of initial polarisation $P_-$ and $P_+$,
respectively. With the same conventions as to upper and
lower signs as in Appendix A, the results are
as follows.

The $\mid \psi_{S^\prime,T^\prime}>$ populations are
\begin{displaymath}
\mid C_{S^\prime,T^\prime}\mid^2=\ \frac{1}{2}
\ (1+x^2)^{-1/2}\ (\sqrt{1+x^2}\pm 1)^{-1}\times
\end{displaymath}
\begin{displaymath}
\times\left\{\ \frac{1}{2}(1+x^2)^{1/2}(\sqrt{1+x^2}\pm 1)
\left[1-P_- P_+\right]\ \mp\right.
\end{displaymath}
\begin{displaymath}
\mp\ \frac{1}{2}(\sqrt{1+x^2}\pm 1)\frac{2\mu}{\Delta E}B_z
\left[P_- -P_+\right]
+\ \left(\frac{2\mu}{\Delta E}\right)^2 (B_- B_+) P_- P_+\  \mp
\end{displaymath}
\begin{equation}
\mp\ \frac{1}{2}(\sqrt{1+x^2}\pm 1)
\sqrt{1-P^2_-}\ \sqrt{1-P^2_+}\ \cos(\phi_- - \phi_+)\ -
\end{equation}
\begin{displaymath}
-\ \frac{1}{4}\left(\frac{2\mu}{\Delta E}\right)^2
\sqrt{1-P^2_-}\ \sqrt{1-P^2_+}
\left[B_- e^{i\phi_+}+B_+ e^{-i\phi_+}\right]
\left[B_- e^{i\phi_-}+B_+ e^{-i\phi_-}\right]\ \pm
\end{displaymath}
\begin{displaymath}
\pm\ \frac{1}{2\sqrt{2}}\left(\frac{2\mu}{\Delta E}\right)
\sqrt{1-P^2_+}\left[B_- e^{i\phi_+} + B_+ e^{-i\phi_+}\right]
\left[\sqrt{1+x^2}\pm 1)\mp\frac{2\mu}{\Delta E}\ B_z\  P_-\right]\ \mp
\end{displaymath}
\begin{displaymath}
\left.\mp\ \frac{1}{2\sqrt{2}}\left(\frac{2\mu}{\Delta E}\right)
\sqrt{1-P^2_-}\left[B_- e^{i\phi_-} + B_+ e^{-i\phi_-}\right]
\left[(\sqrt{1+x^2}\pm 1)\pm\frac{2\mu}{\Delta E}\
B_z\  P_+\right]\right\}\ .
\end{displaymath}

\vspace{10pt}

Similarly, the $\mid \psi_\pm>$ populations are
\begin{displaymath}
\mid C_\pm\mid^2=
\ \frac{1}{4}\frac{B^2_z}{\mid\vec{B}\mid^2}
\left[1+P_- P_+\right]
\pm\ \frac{1}{4}\frac{B_z}{\mid\vec{B}\mid}
\left[P_- + P_+\right]
+\ \frac{1}{2}\frac{B_- B_+}{\mid\vec{B}\mid^2}\ \pm
\end{displaymath}
\begin{displaymath}
\pm\ \frac{1}{4\sqrt{2}}\sqrt{1-P^2_+}
\left[1\pm\frac{B_z}{\mid\vec{B}\mid}P_-\right]
\left[\frac{B_-}{\mid\vec{B}\mid}e^{i\phi_+}+
\frac{B_+}{\mid\vec{B}\mid}e^{-i\phi_+}\right]\ \pm
\end{displaymath}
\begin{equation}
\pm\ \frac{1}{4\sqrt{2}}\sqrt{1-P^2_-}
\left[1\pm\frac{B_z}{\mid\vec{B}\mid}P_+\right]
\left[\frac{B_-}{\mid\vec{B}\mid}e^{i\phi_-}+
\frac{B_+}{\mid\vec{B}\mid}e^{-i\phi_-}\right]\ +
\end{equation}
\begin{displaymath}
+\ \frac{1}{8}\sqrt{1-P^2_-}\ \sqrt{1-P^2_+}
\left[\frac{B_-}{\mid\vec{B}\mid}e^{i\phi_+}+
\frac{B_+}{\mid\vec{B}\mid}e^{-i\phi_+}\right]
\left[\frac{B_-}{\mid\vec{B}\mid}e^{i\phi_-}+
\frac{B_+}{\mid\vec{B}\mid}e^{-i\phi_-}\right]\ .
\end{displaymath}

\vspace{10pt}

Note that in each of these two equations,
the last three terms could be combined further
into the product of two terms. However, results in the form given
here are more amenable to the phase average discussed in the main text.
Incidentally, it is straightforward to check that the sum of all four
populations does indeed reduce to $1$, since, by construction,
the state $\mid \psi, t=0>$
is normalised to $1$.

\end{document}